\documentclass[onecolumn,aps,prd,superscriptaddress,showpacs]{revtex4}

\def\paint#1{#1}

\usepackage{amsmath}
\usepackage{amssymb}
\usepackage{amsfonts}
\usepackage{amsthm}

\def\sym{\textrm{S}}
\def\C{\mathcal{C}}

\def\pt{p_{\,\mathrm{t}}}

\def\d{\mathrm{d}}
\def\im{{\it i}}

\def\fd#1#2{\frac{\delta#1}{\delta#2}}

\def\H{\mathcal{H}}

\def\M{M}
\def\D{D}
\def\P{P}

\def\ang#1#2{\langle#1,#2\rangle}

\def\kuchar{Kucha\v{r}}
\def\A{\mathcal{A}}
\def\B{\mathcal{B}}

\def\nab{\nabla\!_\gamma{}}

\begin{document}

\title{Conformal geometrodynamics: True degrees of freedom in a truly canonical structure}

\author{Charles H.-T. Wang}
\email{c.wang@abdn.ac.uk}
\affiliation{School of Engineering and Physical Sciences,
University of Aberdeen, King's College,
Aberdeen AB24 3UE,
Scotland}
\affiliation{Rutherford Appleton Laboratory, CCLRC, Chilton, Didcot, Oxon OX11 0QX,
England
}


\begin{abstract}
The standard geometrodynamics is transformed into a theory of
conformal geometrodynamics by extending the ADM phase space for
canonical general relativity to that consisting of York's mean
exterior curvature time, conformal three-metric and their momenta.
Accordingly, an additional constraint is introduced, called the
conformal constraint. In terms of the new canonical variables, a
diffeomorphism constraint is derived from the original momentum
constraint. The Hamiltonian constraint then takes a new form. It
turns out to be the sum of an expression that previously appeared
in the literature and extra terms quadratic in the conformal
constraint. The complete set of the conformal, diffeomorphism and
Hamiltonian constraints are shown to be of first class through the
explicit construction of their Poisson brackets. The extended
algebra of constraints has as subalgebras the Dirac algebra for
the deformations and Lie algebra for the conformorphism
transformations of the spatial hypersurface. This is followed by a
discussion of potential implications of the presented theory on
the Dirac constraint quantization of general relativity. An
argument is made to support the use of the York time in
formulating the unitary functional evolution of quantum gravity.
Finally, the prospect of future work is briefly outlined.
\end{abstract}


\maketitle

\section{Introduction}

Finding the correct theory to unify general relativity (GR)  and
quantum mechanics is amongst
the most important outstanding issues in fundamental physics. For
almost a century countless ingenious ideas have been put forward
to this end. The string theory exemplifies those based on a
drastic modification of matter and spacetime structure. There \paint{have}
also been significant recent developments, notably the decoherent
histories theories, that advocate the revision of
the quantization procedure itself. A less radical strategy falls
into the broad category of canonical quantum GR. This strategy
often requires extensive classical analysis in terms of,
e.g. sophisticated canonical transformations of the gravitational
variables, so as to reach an appropriate point of departure for
quantization. To gain advantage, the original phase space for
canonical GR may undergo symplectic reduction or extension by
exploiting symmetry principles. With a minimum \paint{of} alternation to
established physical principles, it offers a natural approach to
quantum gravity and has a long history of investigations.

By leaving aside the conceptional issue of time
and
technical
difficulties in
deparametrizing GR, important progress of canonical quantum GR has been made
on the functional-analytic approach to the gravitational
constraints over the last decade or so.
This is made possible by enlarging the phase space
of GR to allow for a spin gauge symmetry that is amiable
to the background independent ``loop quantization''
\paint{technique}.
The development is also based
on the recognition of various technical issues in implementing
the Dirac constraint quantization. These include the need to
regularize the Wheeler-DeWitt equation, as it involves second
order functional derivatives that may give rise to divergence.
Despite the above encouraging progress, however, the loop quantum gravity
programme has yet a number of major challenges to overcome. Amongst
them are the well known Barbero-Immirzi
ambiguity in choosing the canonical variables and the
persistent lack of preferred time variable to generate
unitary evolution of quantum gravity. While the former
difficulty has recently been \paint{related to} a
{\em conformal} freedom in the Ashtekar type
gauge treatment of the ADM geometrodynamical
variables, the latter has long been suspected to \paint{reflect} the need to
quantize the true \paint{gravitational degrees of freedom}.

In classical
GR, a promising candidate to carry the
true dynamics of the gravitational field
has already been identified
\paint{to be the {\em conformal} three-geometry by York}
over three decades ago \cite{York1971-2}.
Since then,
considerable efforts have been
made to construct a \paint{canonical} approach to
GR in which the
scaling part of kinematics naturally generates the
functional time evolution of the conformal part.
Indeed, one of the most compelling physical reasons for such a construction is
the anticipation
for the transverse-traceless momentum of the
conformal three-geometry to give rise to a spin-2 graviton description
on quantization.
Early works
in this enterprise involve normalizing
the conformal metric to have unit scale factor
locally.  The imposition of this condition
inevitably introduces a nontrivial projection
factor into the Poisson bracket (PB)
relations of the traceless metric momenta \cite{ChoquetBruhatYork1980, Kuchar1992a}.
This leads tantalizingly  to an ``almost Hamiltonian'' formulation of GR,
which is unfortunately not readily amicable to standard canonical quantization.

Recently a coordinate independent normalization scheme
has been considered \cite{FischerMoncrief1997-9} in which the
conformal metric on a slice with
constant mean curvature (CMC) has been rescaled
in the fashion of Yamabe,
so that
the scale curvature ($R$) takes the value of $\pm1$ or $0$.
It has been successfully demonstrated that at least in the case of
hyperbolic compact three-spaces with $R=-1$, a full classical Hamiltonian reduction
of GR can be accomplished. Formally, the reduced Hamiltonian density is
simply the scale factor. However it is implicitly defined to satisfy
a nonlinear elliptic equation and therefore is hard to be turned into
a quantum operator. Besides, it is not clear whether
general covariance could be retained if GR is to be quantized in
a formalism relying on a privileged foliation.

In this paper
we present  how a genuine  canonical
evolution of conformal three-geometry for arbitrary spacetime foliations
can be
achieved by explicit construction of a new form of
Hamiltonian for GR.
By judicious use
of Dirac's theory of first
class constraints,
we show how
various concerns over
previous approaches described above are transcended.
Although the main results of this paper lie in the classical domain,
it provides the necessary canonical analysis towards
a new direction for loop quantum gravity by incorporating
conformal symmetry as well as spin symmetry. Such a direction may \paint{lead to} fresh
insight and new tools in addressing
the problems of time, unitary evolution and Barbero-Immirzi ambiguity
in {\em quantum general relativity}.

\section{Conventional geometrodynamics}

To establish convention,
we start by recapitulating the standard ADM
paradigm of canonical gravity \cite{DeWitt1967}
with metric signature $(-,+,+,+)$ and compact spatial sectors.
The three-metric, its inverse, scale factor (unit density function)
and associated Levi-Civita connection
are
respectively denoted by $g_{a b}$, $g^{ab}$, $\mu := \sqrt{\det g_{ab}}$ and
$\nabla$, for $a, b, \cdots = 1,2,3$. The extrinsic curvature is
given by
\begin{equation}\label{Kab}
K_{a b} = \frac1{2N}\left(\dot{g}_{a b} - \nabla_a N_b - \nabla_b N_a  \right)
\end{equation}
where $N$ is
the lapse function  and $N^a$
the shift vector. In terms of \eqref{Kab} and the mean curvature
$K := g^{ab} K_{ab}$, the canonical momentum of $g_{ab}$
is given by
\begin{equation}\label{}
p^{ab} = \mu (K^{ab}-g^{ab} K  ) .
\end{equation}
Note that, although for convenience, our sign convention for the exterior curvature
differs from that of \cite{DeWitt1967}, the expression
of $p^{ab}$ in terms of the metric and its derivatives is the same.
We denote by $G$ the ADM phase space
for canonical GR with coordinates \paint{$(g_{ab}, p^{ab})$}.
These variables will also be referred to as $G$-variables.
The canonical form of Einstein's equation is generated by
the well established ADM action given by
\begin{equation}\label{SDADM}
S_G
=
\int
\!
\left(
p \cdot \dot{g} - H_G
\right)\d t
\end{equation}
where $p \cdot \dot{g} := \int {p}^{ab} \dot{g}_{ab}\,\d^3 x$
with the over dot denoting a $t$-derivative,
and
\begin{equation}\label{H0}
H_G :=
\int\!
\left(
N \H + N^a \H_a
\right)\d^3 x .
\end{equation}
This ``Hamiltonian'' consists of the
momentum constraint:
\begin{equation}
\label{Jconstr}
\H_a
=
-2\, \nabla_b \,{p}^b{}_a
\end{equation}
and Hamiltonian constraint:
\begin{equation}
\label{Hconstr}
\H
=
\mu^{-1} g_{a b c d}\, p^{a b} p^{c d} -  \mu R
\end{equation}
were $R$ is the (Ricci) scalar curvature of $g_{ab}$ and
\begin{equation}\label{}
g_{a b c d} := \frac1{2} \left(g_{a c}g_{b d} + g_{a d}g_{b c} -  g_{a b}g_{c d} \right).
\end{equation}

The Poisson bracket for any functionals $\A$ and $\B$
with respect to the $G$-variables
$(g_{ab}, p^{ab})$
is defined by
\begin{equation}\label{AABB}
\{\A, \B \}_G := \int \left[
\fd{\A}{{g}_{a b}(x)}\fd{\B}{{p}^{a b}(x)}
-
\fd{\B}{{g}_{a b}(x)}\fd{\A}{{p}^{a b}(x)}
\right] \d^3  x .
\end{equation}
The canonical PB relations follows as:
\begin{align}
\label{CPB1}
\{ g_{a b}(x), p^{c d}(x') \}_\Gamma  &= \delta^{c d}_{a b}\, \delta(x, x')\\
\label{CPB2}
\{ g_{a b}(x),  g_{c d}(x') \}_\Gamma &= \{ p^{a b}(x), p^{c d}(x') \}_\Gamma = 0.
\end{align}
The PBs of the momentum and Hamiltonian constraints
satisfy the well known Dirac algebra \cite{DeWitt1967, Dirac1964}:
\begin{align}
\label{HiHj}
\hspace{-2pt}
\{\H_a(x), \H_b(x')\}_G
&=
\H_b(x)\,\delta_{,a}(x, x')
-
(a x\leftrightarrow b x')
\\
\label{H0H0}
\{\H(x), \H(x')\}_G
&=
(g^{ab}\H_a)(x)\,\delta_{,b}(x, x')
-
(x\leftrightarrow x')
\\
\label{HiH0}
\{\H_a(x), \H(x')\}_G
&=
 \H(x) \,\delta_{,a}(x, x') .
\end{align}
The structure of this algebra guarantees that the evolution of
the intrinsic and extrinsic
three-geometry using the canonical equations of motion
\begin{align}
\label{ceq1}
\dot{g}_{ab} &= \{{g}_{ab}, H_G\}_G\\
\label{ceq2}
\dot{p}^{ab} &= \{{p}^{ab}, H_G\}_G
\end{align}
subject to the constraint equations
\begin{align}
\label{admHa=0}
\H_a &= 0\\
\label{admH=0}
\H &= 0
\end{align}
is independent of how the spatial
hypersurface is deformed and coordinatized
from
the initial to final configurations
for arbitrary $N(x^a,t)$ and $N^a(x^a,t)$
\cite{HojmanKucharTeitelboim1973}.

\section{Extended phase space of GR with conformal symmetry}

We shall now reformulate
the canonical dynamics of GR to explicitly depend on
the underlying conformal spatial metric.
This is  motivated by previous works in the
conformal treatment of GR
(see e.g.
\cite{York1971-2, York1973-4, FischerMarsden1977, IsenbergOMurchadhaYork1976, ChoquetBruhatYork1980, Kuchar1992a, FischerMoncrief1997-9})
as well as recent efforts in
addressing the unitary evolution and
the conformal
approach to quantum gravity
\cite{Wang}.
The present approach features a
number of new perspectives and generalities:
\paint{(i)} it is free from the constant mean curvature (CMC) condition,
\paint{(ii)} no normalization condition on the conformal metric will be imposed, and
\paint{(iii)} instead, Dirac's theory of first class constraints will be employed
to explore and exploit the conformal symmetry of an extended phase
space for canonical GR.

Consider the decomposition of the metric momentum ${p}^{ab}$
into its trace-free part and trace part according to
\begin{equation}\label{psplit}
p^{ab}
=
\pt^{ab} + \frac13  \, p\, g^{ab}
\end{equation}
where $p := g_{ab} {p}^{ab} = -2\,\mu K$
such that $g_{ab} \pt^{ab} = 0$. Conformal transformations may be
applied to the two irreducible pieces of ${p}^{ab}$ in
\eqref{psplit}. The metric ${g}_{ab}$ and traceless part of the
metric momentum $\pt^{ab}$ are then conformally related to a
symmetric tensor $\gamma_{ab}$ and symmetric tensor density
$\pi^{ab}$ respectively as follows:
\begin{equation}\label{ggamma}
{g}_{ab} = \phi^4 \gamma_{ab}
\end{equation}
\begin{equation}\label{pt}
\pt^{ab} = \phi^{-4} \pi^{ab}
\end{equation}
using a positive conformal factor \paint{$\phi$}. The trace-free
condition of $\pt^{ab}$ implies a similar condition of
$\pi^{ab}$ with respect to $\gamma_{ab}$. However, rather than
restricting the independent components of
$\pi^{ab}$ from the outset,
we choose to
implement the trace-free condition of $\pi^{ab}$ ``weakly'' {\it a la} Dirac \cite{Dirac1950}
by introducing a new constraint
\begin{equation}\label{conformconst}
\C
:=
{\gamma}_{ab}\pi^{ab}
\end{equation}
which will only be required to vanish alongside the Hamiltonian and momentum constraints.
The advantage of this approach will become evident below.
The tensor $\gamma_{ab}$ will play the role of the conformal metric,
whereas the
original $g_{ab}$ will stay as the physical metric.
Introducing
the conformal scale factor
\begin{equation}\label{}
\mu_\gamma := \sqrt{\det \gamma_{ab}}
\end{equation}
we see from
\eqref{ggamma} that
$\phi$ may be expressed as a function of
$\gamma_{ab}$ and $\mu$ given by
\begin{equation}\label{phi}
\phi = \phi(\gamma_{ab}, \mu) = \left(\frac{\mu}{\mu_\gamma}\right)^{1/6}.
\end{equation}
Using  \eqref{pt}
we can rewrite \eqref{psplit} as
\begin{equation}\label{pexpr}
p^{ab}
=
\phi^{-4} \pi^{ab} - \frac12  \, \phi^2 \mu_\gamma\, \gamma^{ab}\tau
\end{equation}
where
$\tau:=\frac{4}{3}\, K$ which
will be designated as the ``York time''.
Note that
all components of the symmetric tensors ${\gamma}_{ab}$ and $\pi^{ab}$
are treated as independent.

By virtue of the expression \eqref{pexpr} we may calculate the time
derivative terms in the ADM action \eqref{SDADM}, resulting in
\begin{equation}\label{pdotg}
p^{ab} \dot{g}_{ab}
=
-\tau \, \dot{\mu}
+
\pi^{ab} \dot{\gamma}_{ab}
+4(\ln\phi)\,\dot{}\, \C .
\end{equation}
This way
the pairs  $({\gamma}_{ab}, \pi^{ab})$ and $(\tau, \mu)$
emerge as new canonical variables in place of
$(g_{ab}, p^{ab})$ so long as $\C$ can be consistently treated as a
constraint.
To see that this is indeed the case, we further show that
the momentum and Hamiltonian constraints now take the following forms:
\begin{align}\label{diffconst}
\H_a
&=
\tau_{,a} {\mu} -2\, \nab_b \,\pi^b{}_a
+4(\ln\phi)_{,a}\, \C
\\
\label{HR}
\H
&=
-\frac38\, \tau^2 \mu
+
\frac1{\mu}\,\pi{}_{ab}\pi^{ab}
-\mu R
+\frac\tau2\,\C
-\frac1{2\mu}\,\C^2 .
\end{align}
Here both $\C$ and  $\phi$ in turn \paint{depend} on the new variables
as specified in \eqref{conformconst}
and \eqref{phi} respectively.
The appearance of $\C$
as subexpressions of $\H_a$ and $\H$
above is a necessary consequence the ``weak''
treatment of this constraint.
The indices of $\pi^{ab}$ and $\pi_{ab}$
are raised or lowed by the conformal metric $\gamma_{ab}$
and its inverse $\gamma^{ab}$ respectively.
The scalar curvature $R$ of $g_{ab}$
in \eqref{HR}
may be expressed
in terms of $\mu$ and $\gamma_{ab}$ as
\begin{equation}\label{Rg}
R
=
\phi^{-4} R_\gamma -8\,\phi^{-5} \Delta_\gamma \phi
\end{equation}
using the Levi-Civita connection $\nab$,
scalar curvature $R_\gamma$ and
Laplacian $\Delta_\gamma := \gamma^{ab}\nab_a\nab_b$,
associated with the conformal metric $\gamma_{ab}$.

Based on the above expressions,
we can  construct an action integral for GR
equivalent to \eqref{SDADM} as follows:
\begin{equation}\label{SDADM1}
S_\Gamma
=
\int
\!
\left(
\mu \cdot \dot{\tau}
+
\pi \cdot \dot{\gamma} - H_\Gamma
\right)\d t
\end{equation}
where
\eqref{pdotg} has been used
with a total time derivative dropped. The Hamiltonian above is given by
\begin{equation}\label{H1}
H_\Gamma :=
\int\!
\left(
N \H + N^a \H_a + \Phi \,\C
\right)\d^3 x
\end{equation}
with an additional term
$\Phi \,\C$ where $\Phi=\Phi(x,t)$ is a new Lagrange multiplier
used to effect weakly vanishing of $\C$.
The presence of this additional constraint offsets the
redundancy of $\gamma_{ab}$ and $\pi^{ab}$
due to the fact that they are unique up to a local rescaling.
We denote by
$\Gamma$ the extended phase space, having the new
\paint{``conformo-geometrodynamical variables''}
$(\gamma_{ab}, \pi^{ab}; \tau, \mu)$ as coordinates. They will also be
referred to as the $\Gamma$-variables.


\section{Structure of the new constraints and the new Hamiltonian}

To determine the main mathematical structure
and physical interpretation of the enlarged set of
constraints including $\C$, it is necessary to
compute the PBs of various
quantities involved in the analysis with respect to
the $\Gamma$-variables
$(\gamma_{ab}, \pi^{ab}; \tau, \mu)$ given by
\begin{equation}\label{PBAB}
\{ \A, \B\}_\Gamma :=
\int \left[
\fd{\A}{{\gamma}_{a b}(x)}\fd{\B}{{\pi}^{a b}(x)}
-
\fd{\B}{{\gamma}_{a b}(x)}\fd{\A}{{\pi}^{a b}(x)}
+
\fd{\A}{\tau(x)}\fd{\B}{\mu(x)}
-
\fd{\B}{\tau(x)}\fd{\A}{\mu(x)}
\right]\d^3  x .
\end{equation}
The corresponding canonical PB relations follow as:
\begin{align}
\label{}
\{ \gamma_{a b}(x), \pi^{c d}(x') \}_\Gamma  &= \delta^{c d}_{a b}\, \delta(x, x')\\
\label{}
\{ \tau(x), \mu(x') \}_\Gamma  &=  \delta(x, x')\\
\label{}
\{ \gamma_{a b}(x),  \gamma_{c d}(x') \}_\Gamma &= \{ \pi^{a b}(x), \pi^{c d}(x') \}_\Gamma = 0 \\
\label{}
\{ \tau(x),  \tau(x') \}_\Gamma &= \{ \mu^{a b}(x), \mu^{c d}(x') \}_\Gamma = 0
\end{align}

For brevity we shall abbreviate $\{ \A, \B\}_\Gamma$
by $\{ \A, \B\}$
in the following discussion.
Using \eqref{ggamma}, \eqref{pexpr} and the variational identity
\begin{equation}\label{varphi}
\delta \phi
=
\frac 1 6 \,\frac{\phi}{\mu}\, \delta\mu
-
\frac 1 {12}\,\phi \,\gamma^{ab} \delta \gamma_{ab}
\end{equation}
we see the following PB relations
\begin{align}
\label{CCPB1}
\{ g_{a b}(x), p^{c d}(x') \}  &= \delta^{c d}_{a b}\, \delta(x, x')\\
\label{CCPB2}
\{ g_{a b}(x),  g_{c d}(x') \} &= \{ p^{a b}(x), p^{c d}(x') \} = 0
\end{align}
inherited from the PB relations \eqref{CPB1} and \eqref{CPB2}
for the
original geometrodynamical variables.
An important consequence of this is that
the Dirac algebra in \eqref{HiHj}--\eqref{HiH0}
is \paint{``strongly''} preserved by regarding
all quantities involved as functions of
the $\Gamma$-variables rather than the $G$-variables.
For instance, the factor $g^{ab}$ on the RHS of \eqref{H0H0} is given by
\begin{equation}\label{gabup}
g^{ab} = \phi^{-4}\gamma^{ab} = \left(\frac{\mu_\gamma}{\mu}\right)^{2/3} \!\gamma^{ab}
\end{equation}
as per \eqref{ggamma} and \eqref{phi}.
A similar statement can equally be made for the
PBs of any functionals depending on
$({\gamma}_{ab}, \pi^{ab}; \tau, \mu)$ through
$(g_{a b}, p^{ab})$.
Straight form the definition of $\C$ we see that
\begin{align}\label{PBCC}
\{\C(x),  \C(x')\} &= 0 \\
\label{PB_gamma_C}
\{ \gamma_{ab}(x), \C(x')\} &=  \gamma_{ab}(x)\, \delta(x,x') \\
\label{PB_pi_C}
\{ \pi^{ab}(x), \C(x')\} &= -\pi^{ab}(x)\, \delta(x,x')
\end{align}
which reveal
$\C$ as the canonical generator of the conformal transformations. Accordingly,
$\C$ will be referred to as the ``conformal constraint''.
This interpretation is consistent with
$\tau$ and $\mu$ being conformally invariant, i.e.
\begin{equation}\label{PBCtaumu}
\{ \tau(x), \C(x')\} = 0,\quad
\{ \mu(x), \C(x')\} = 0 .
\end{equation}
By virtue of \eqref{PB_gamma_C}, \eqref{PB_pi_C}, \eqref{PBCtaumu} and
\begin{equation}\label{PBCphi}
\{\phi(x), \C(x')\} = -\frac14\,\phi(x)\, \delta(x,x')
\end{equation}
derived from \eqref{varphi} we can
reaffirm that
the physical metric $g_{a b}$ and momentum
$p^{a b}$
are themselves conformally invariant, i.e.
\begin{equation}\label{PBCgp}
\{ \C(x), g_{a b}(x')\} = 0,\;
\{ \C(x),  p^{a b}(x')\} =0 .
\end{equation}
Eq. \eqref{PBCphi} implies that the quantity $Q$ defined by
$\phi^4 = e^{-Q}$
is canonically conjugate to $\C$ since
\begin{equation}\label{PB_Q_C}
\{-4\ln(\phi(x)), \C(x')\} = \delta(x,x') .
\end{equation}
It is useful to observe that
\begin{align}
\{\phi(x), \phi(x')\} &= 0\\
\{\phi(x), g_{a b}(x')\} &= 0\\
\{\phi(x), p^{a b}(x')\} &= 0
\end{align}
by using the  identities
\begin{align}
\label{}
\{\mu(x), p^{a b}(x')\}
&=  \frac12\,(\phi^2 \mu_\gamma\gamma^{a b})(x)\, \delta(x,x') \\
\label{}
\{\mu_\gamma(x), p^{a b}(x')\}
&= \frac12\,(\phi^{-4} \mu_\gamma\gamma^{a b})(x)\, \delta(x,x') .
\end{align}
Hence, as well as $\C$,
the PBs of $\phi$ with any functionals depending on
and $({\gamma}_{ab}, \pi^{ab}; \tau, \mu)$
through $(g_{a b}, p^{a b})$
vanish strongly. From this, a host of mathematical niceties can be derived.
Of particular interest in the present analysis are the
vanishing PBs for the Hamiltonian and
momentum constraints with $\C$ and $\phi$:
\begin{align}
\label{PBCHP}
\{ \H(x),  \C(x')\} &= \{ \H_a(x),  \C(x')\} = 0 \\
\label{PB_H_phi}
\{ \H(x),  \phi(x')\} &= \{ \H_a(x),  \phi(x')\} = 0 .
\end{align}
Consequently, the expression $4(\ln\phi)_{,a}\, \C$ as the last term of $\H_a$ in \eqref{diffconst}
has zero PBs with the Hamiltonian and
momentum constraints. Therefore, if we introduce the ``diffeomorphism constraint'':
\begin{equation}\label{diffconst1}
\C_a
:=
\H_a
-
4(\ln\phi)_{,a}\, \C
=
\tau_{,a}{\mu} -2\, \nab_b \,\pi^b{}_a
\end{equation}
then it follows from the Dirac algebra, \eqref{PBCC},
\eqref{PBCphi}, \eqref{PBCHP} and \eqref{PB_H_phi} that
the complete set of independent constraints $\{\C,\C_a,\H\}$ is first class
satisfying the following algebra:
\begin{align}
\label{PBCC1}
\{\C(x),  \C(x')\}
&=
0
\\
\label{CiCjgamma}
\{\C_a(x), \C_b(x') \}
&=
\C_b(x)\,\delta_{,a}(x, x')
-
(a x\leftrightarrow b x')
\\
\label{PBCCa}
\{\C_a(x),  \C_(x')\}
&=
\C(x)\, \delta_{,a}(x, x')
\\
\label{HHgamma}
\{\H(x), \H(x') \}
&=
\phi^{-4}\gamma^{ab} (\C_a + 4(\ln\phi)_{,a}\, \C)(x)\,\delta_{,b}(x, x')
-
(x\leftrightarrow x')
\\
\label{PBCH}
\{ \C(x),  \H(x')\}
&=
0
\\
\label{CiHgamma}
\{\C_a(x), \H(x') \}
&=
 \H(x)\, \delta_{,a}(x, x')
\end{align}
where $\phi$
depends on $\gamma_{ab}$ and $\mu$ according to
$\phi = ({\mu}/{\mu_\gamma})^{1/6}$
as specified in \eqref{phi}.

It is clear that $\C$ and $\C_a$
are the canonical generators of the
conformal  and diffeomorphism transformations,
with the corresponding groups denoted by $\P$ and $\D$ respectively.
Furthermore, the subset
consisting of
\eqref{PBCC1}, \eqref{CiCjgamma} and \eqref{PBCCa} is a manifestation of
the Lie algebra of the {\em conformorphism} group $C$
as the semi-direct product of $\P$ and $\D$. Such a group
structure goes back to the original
study of the initial value problem in
GR
\cite{York1973-4, FischerMarsden1977}.

In terms of the constraints $\C$, $\C_a$ and $\H$,
we are finally led to the canonical action and its Hamiltonian of the form

\begin{equation}\label{Sfinal}
S
=
\int
\!
\left(
\mu \cdot \dot{\tau}
+
\pi \cdot \dot{\gamma} - H
\right)\d t
\end{equation}
\paint{and}
\begin{equation}\label{Hfinal}
H :=
\int\!
\left(
N \H + N^a \C_a + \Phi \,\C
\right)\d^3 x
\end{equation}
which is equivalent to \eqref{H1} with a slightly  redefined $\Phi$ due to \eqref{diffconst1}.

This Hamiltonian generates the following
canonical equations of motion
\begin{align}
\label{ceqfinal1}
\dot\tau &= \{\tau, H \}\\
\label{ceqfinal2}
\dot\mu &= \{\mu, H \}\\
\label{ceqfinal3}
\dot\gamma_{ab} &= \{\gamma_{ab}, H \}\\
\label{ceqfinal4}
\dot\pi^{ab} &= \{\pi^{ab}, H \}
\end{align}
subject to the
conformal, diffeomorphism and Hamiltonian
constraint equations:
\begin{align}
\label{C=0}
\C &= 0\\
\label{Ca=0}
\C_a &= 0\\
\label{H=0}
\H &= 0
\end{align}
respectively.
The dynamical consistency of the these
equations is guaranteed by
the first class nature of the constraints
$\C$, $\C_a$ and $\H$.

\section{Hamiltonian reduction to the ADM formalism of GR}


\paint{We shall} demonstrate that
the new system of equations
\eqref{ceqfinal1}--\eqref{ceqfinal4}
and
\eqref{C=0}--\eqref{H=0}
\paint{does indeed imply}
the
original ADM system of
equations for GR, i.e.
\eqref{ceq1}, \eqref{ceq2},
\eqref{admHa=0} and \eqref{admH=0}.
Equations \eqref{ggamma}
and \eqref{pexpr} augmented by \eqref{phi}
enable the mapping of solutions in the
$\Gamma$-variables to equivalent solutions
in the $G$-variables.

To show that
the ADM canonical equations of motion 
\paint{are satisfied by}
the recovered metric and its momentum we evaluate
the time derivative of ${g}_{ab}$ on
the constraint surface where $\C=0$ as follows:
\begin{align}
\dot{g}_{ab}(x)
&=
\int\left[
\fd{{g}_{ab}(x)}{{\gamma}_{cd}(x')}\,\dot{\gamma}_{cd}(x')
+
\fd{{g}_{ab}(x)}{{\pi}^{cd}(x')}\,\dot{\pi}^{cd}(x')
+
\fd{{g}_{ab}(x)}{\tau(x')}\,\dot{\tau}(x')
+
\fd{{g}_{ab}(x)}{\mu(x')}\,\dot{\mu}(x')
\right]\d^3x'
\label{step10}\\
&=
\int\left[
\fd{{g}_{ab}(x)}{{\gamma}_{cd}(x')}\,\fd{H}{{\pi}^{cd}(x')}
-
\fd{{g}_{ab}(x)}{{\pi}^{cd}(x')}\,\fd{H}{{\gamma}_{cd}(x')}
+
\fd{{g}_{ab}(x)}{\tau(x')}\,\fd{H}{\mu(x')}
-
\fd{{g}_{ab}(x)}{\mu(x')}\,\fd{H}{\tau(x')}
\right]\d^3x'
\label{step20}
\end{align}
where the new canonical equations of motion \eqref{ceqfinal1}--\eqref{ceqfinal4} have been used.

Note that on
the constraint surface of $\C$ we have the following
relation
\begin{align}
\fd{H}{{\pi}^{cd}(x')}
&=
\int\left[
\fd{{g}_{ef}(x'')}{{\pi}^{cd}(x')}\,\fd{H_G}{{g}_{ef}(x'')}
+
\fd{{p}^{ef}(x'')}{{\pi}^{cd}(x')}\,\fd{H_G}{{p}^{ef}(x'')}
\right]\d^3x''
\end{align}
as well as similar relations obtained by replacing ${\pi}^{cd}(x')$ above
with ${\gamma}_{cd}(x')$, ${\tau}(x')$ and  ${\mu}(x')$ respectively.
Substituting these relations into \eqref{step20}
we \paint{see that}
\begin{align}
\dot{g}_{ab}(x)
&=
\int\left[
\{{g}_{ab}(x), {g}_{ef}(x')\}\,\fd{H_G}{{g}_{ef}(x')}
+
\{{g}_{ab}(x), {p}^{ef}(x')\}\,\fd{H_G}{{p}^{ef}(x')}
\right]\d^3x'
\nonumber\\
&=
\fd{H_G}{{p}^{ab}(x)} = \{{g}_{ab}(x), H_G\}_G
\label{dotgok}
\end{align}
where the PB relations \eqref{CCPB1} and \eqref{CCPB2} have been used.
The above result is identical to \eqref{ceq1}.

Following arguments similar to those leading from \eqref{step10} to \eqref{dotgok}
we can show that
\begin{align}
\dot{p}^{ab}(x)
&=
-\fd{H_G}{{g}_{ab}(x)} = \{{p}^{ab}(x), H_G\}_G
\label{dotpok}
\end{align}
which is identical to \eqref{ceq2}.

By taking into account \eqref{diffconst1}
we see immediately that if the new constraint equations
\eqref{C=0}--\eqref{H=0} are satisfied,
the ADM constraint equations
\eqref{admHa=0} and \eqref{admH=0} are
satisfied as well.
We have therefore recovered
the ADM description of GR as expected.

\section{Discussion on quantization issues and outlook}

Having reformulated the conventional geometrodynamics to conformal
geometrodynamics, we shall be interested in whether the new
formalism of canonical GR bears any implication on
the quantum gravity programme. Would this approach to the York
time for arbitrary spacetime foliations shed new light on the
problem of time \cite{Isham1993} and hence on the issue of unitary
evolution in quantum gravity?
Let us briefly address these issues using somewhat heuristic
arguments with a view to more complete analysis.

We will for a moment
work in the $(\gamma_{ab},\tau)$ representation and aim to treat
$\tau$ as the functional time. Let $\M$ be the space of
Riemannian metrics with $\gamma_{ab}$ as elements.
Due to the presence of the conformorphism symmetry,
the effective configuration space under consideration is the
quotient $\M/\P/\D$, i.e. the (quantum) conformal superspace
\cite{York1973-4, FischerMoncrief1996b}.
Consider the state functional $\Psi =  \Psi[\gamma_{ab},\tau]$ and
adopt the formal functional integral
\begin{equation}\label{inn}
\ang{\Psi_1}{\Psi_2} := \int_{\M/\P/\D}\!\!\Psi_1^*\Psi_2\,\delta\gamma
\end{equation}
as the  inner product of two state functionals $\Psi_1$ and $\Psi_2$.
In proceeding with the Dirac constraint quantization scheme,
we will omit the ``hat'' over operators converted from
classical variables for brevity.
Specifically
$\mu = -\im\fd{}{\tau}$ and $\pi^{ab} = -\im\fd{}{\gamma_{ab}}$ should be understood.
With this in mind,
the constraints \eqref{C=0}, \eqref{Ca=0} and \eqref{H=0} are formally turned into the following
quantum constraint equations:
\begin{align}
\label{QC=0}
\C\Psi&=0\\
\label{QCa=0}
\C_a\Psi&=0\\
\label{QH=0}
\H\Psi&=0
\end{align}
respectively.
Concerning factor ordering, we shall pay some attention to the following choice:
\begin{align}\label{OPC}
\C
&=
\sym\left(
{\gamma}_{ab}\pi^{ab}
\right)
\\
\label{OPCa}
\C_a
&=
\tau_{,a}{\mu}
+
\sym\left(
-2\, \nab_b \,\pi^b{}_a
\right)
\\
\label{OPH}
\H
&=
-\frac38\, \tau^2 \mu
+
\sym\left(
\frac1\mu\,\pi{}_{ab}\pi^{ab}
-\mu R
+\frac\tau2\,\C
-\frac{\C^2}{2\mu}
\right)
\end{align}
where
$\sym$ denotes symmetrization with respect to $\gamma_{ab}$ and $\pi^{ab}$.
Thus the term $\sym(\pi{}_{ab}\pi^{ab})$ may naturally be
chosen to be the Laplacian of the conformal superspace.
Note that the leading terms in \eqref{OPCa} and \eqref{OPH} have $\mu$ consistently standing
to the {\em right}. The usefulness of this is as follows:
Take a ``physical'' state functional $\Psi$ satisfying
\eqref{QC=0}--\eqref{QH=0}.
Owing to the described factor ordering,
the Hermiticity
(with respect to the inner product \eqref{inn})
of the second term on the RHS of
\eqref{OPCa} alone is sufficient for one
to conclude that
\begin{equation}\label{dinn}
\ang{\Psi}{\im\fd{\Psi}{\tau(x)}} = \ang{\im\fd{\Psi}{\tau(x)}}{\Psi}
\end{equation}
so long as it is evaluated at a space point $(x^a)$ with
nonzero gradient of $\tau(x)$, i.e.
\begin{equation}\label{nzgrad}
\tau_{,a}(x) \neq 0 .
\end{equation}

Take now a {\em strictly non-CMC} foliation of spacetime
so that nowhere $\tau_{,a}(x)$ vanishes. Using \eqref{dinn}
and the relation
$\dot\Psi=\int \dot\tau(x)\,\fd{\Psi}{\tau(x)}\,\d^3 x$, a
straightforward formal calculation shows that
$\ang{\Psi}{\Psi}\,\dot{}=0$
thereby implying the conservation of the ``total probability''.
For any canonical transformation of the form
$(\gamma_{ab}, \pi^{ab}; \tau, \mu)\rightarrow(X^A,P_A; \tau, \mu)$,
the above discussion can in principle be repeated in the
$(X^A,\tau)$ representation.

It would be desirable to fully justify \eqref{dinn} without
evoking the caveat \eqref{nzgrad}. This will form a subject for
further investigation. Nevertheless, our present discussion
strongly suggests that an implementation of the Dirac quantization
of GR using the described conformal formalism may
lead to the unitary description of quantum  gravity. A particular
strategy along this line currently pursued by the author is to
assimilate the connection approach into the conformal
framework, with an aim to construct a theory of ``conformal loop
quantum gravity''.
The progress of this work will be reported
elsewhere \cite{Wang2005x}.


\begin{acknowledgments}

\paint{I am most grateful to Arthur E. Fischer  and  Chris J. Isham for
stimulating conversations and helpful correspondence.
The research was begun during my previous appointment
at Lancaster University and is
partially supported by the CCLRC Centre for
Fundamental Physics.}

\end{acknowledgments}

\end{document}